\documentclass[aps,pra,twocolumn]{revtex4}   
\newcommand {\be}{\begin{equation}}
\newcommand {\ee}{\end{equation}}
\newcommand {\bey}{\begin{eqnarray}}
\newcommand {\eey}{\end{eqnarray}}
\usepackage{epsfig}
\begin{document}

\title{Impurity Scattering in a Bose-Einstein Condensate at finite
temperature.}

\author{Alberto Montina}
\affiliation{Istituto Nazionale di Ottica Applicata, Largo E.~Fermi 6,50125 Firenze,
Italy}
\date{\today}

\begin{abstract}
We consider the effects of finite temperature on the scattering of impurity atoms in
a Bose-Einstein condensate, showing that the scattering rate is enhanced by the 
thermal atoms. Collisions can increase or decrease the impurity energy.
Below the Landau velocity only the first process occurs, i.e., the collisions cool the
condensate. Above the critical velocity the dissipative collisions prevail over the 
cooling ones for sufficiently low temperatures. These considerations are applied 
to a recent experiment. 
\end{abstract}

\maketitle

Superfluidity is a phenomenon observed for the first time in $^4He$.
It consists in the suppression of the viscosity of the superfluid below a critical
velocity $v_L$, due to linear 
behavior of the dispersion curve $E(p)$ in the neighborhood of $p=0$. 
Landau~\cite{landau} showed with kinematic arguments that $v_L=min[E(p)/p]$. 
Superfluidity can be observed also in atomic Bose-Einstein condensates (BEC),
Bogoliubov equations indicate that $v_L$ is equal to the local speed
of sound. 
The first evidence of a critical velocity in a BEC was obtained by stirring the
condensate with a laser beam~\cite{raman}, but the observed critical velocity was
much lower than the speed of sound, because of the presence of quantized 
vortices~\cite{matthews,madison}.

A recent experiment \cite{ketterle} demonstrated
the superfluid suppression of scattering in agreement with the Landau criterion.
This outcome was obtained by propagating microscopic impurities through the condensate.
The impurities were created by a Raman transition from the $|F=1,m_F=-1\rangle$ state
to the untrapped $|F=1,m_F=0\rangle$ state. The velocity acquired by the untrapped
atoms was due mainly to gravitational acceleration. The radial trapping frequency was
varied to change the velocity of sound and the average impurity velocity.
The number of the overall scattered atoms was evaluated for some values of the radial frequency.
A suppression of the scattering was observed when the average velocity was near the
Landau velocity at the center of the condensate. In Ref.~\onlinecite{ketterle}
a zero-temperature theory is used to evaluate the scattering rate~\cite{nota}. 

In this work we first evaluate the scattering rate at finite temperature
for an homogeneous system,
demonstrating that the thermal cloud can enhance the scattering rate. This effect
would provide an indirect measurement of the condensate temperature. 
We then evaluate the number of scattered 
atoms in a trapped system and show that the thermal atoms may have an observable
effect upon the experimental data of Ref.~\onlinecite{ketterle}.
Furthermore, we calculate the variation of temperature per impurity atom, showing 
that the impurity collisions cool the condensate for tight confinements.

The scattering rate at $T=0$ can be evaluated using Fermi's golden 
rule~\cite{pines,ketterle},
\bey
\nonumber
\Gamma=n_0\left(\frac{2\hbar a}{M}\right)^2\int dq d\Omega q^2 S(q) \\
\times\delta
\left(\frac{\hbar \vec k\cdot\vec q}{M}-\frac{\hbar q^2}{2M}-\omega_q^B\right),
\eey
where $n_0$ is the condensate density, $M$ the atomic mass, $a$ the scattering length for s
wave collisions 
between the impurity atoms and the condensate atoms, and $S(q)=\omega_q^0/\omega_q^B$
the static structure factor of the condensate, with $\hbar\omega_q^0=\hbar^2q^2/2M$ and
$\hbar\omega_q^B=\sqrt{\hbar\omega_q^0(\hbar\omega_q^0+2\mu)}$ being the energies of
a free particle and a Bogoliubov quasiparticle of momentum $q$, respectively. $\mu$ is
the chemical potential.

In the scattering process
an impurity atom changes its momentum and a phonon in the
condensate is created. If the temperature is finite and the phononic modes are 
already populated we expect that the scattering is enhanced because of the bosonic
stimulation, i.e.,

\bey
\nonumber
\Gamma_1(\beta)=n_0\left(\frac{2\hbar a}{M}\right)^2\int dq d\Omega q^2 S(q) \\
\times\delta
\left(\frac{\hbar \vec k\cdot\vec q}{M}-\frac{\hbar q^2}{2M}-\omega_q^B\right)
\frac{e^{\beta\hbar\omega_q^B}}{e^{\beta\hbar\omega_q^B}-1},
\eey
where $1/\beta\equiv T$ is the condensate temperature.

At finite temperatures there is another important channel that contributes to $\Gamma$:
in a scattering process an impurity can change its momentum {\it annihilating} a phonon.
In this case a thermal atom loses energy and increases the condensate population
(this picture is exact for non-interacting atoms, but it is not completely appropriate 
for interacting ones), therefore the
second channel is important because a single quantum state is macroscopically
populated. We get the following additional contribute ({\it cooling scattering rate})
\bey
\nonumber
\Gamma_2(\beta)=n_0\left(\frac{2\hbar a}{M}\right)^2\int dq d\Omega q^2 S(q) \\
\times\delta
\left(-\frac{\hbar \vec k\cdot\vec q}{M}-\frac{\hbar q^2}{2M}+\omega_q^B\right)
\frac{1}{e^{\beta\hbar\omega_q^B}-1}.
\eey
There are two differences between $\Gamma_1$ and $\Gamma_2$. First, in the Dirac's delta the 
sign of $\omega_q^B$ and $\vec k$ are changed, because a phonon is annihilated in the second 
channel $(II)$, whereas it is created in the first one $(I)$.
The second difference is the additional temperature-dependent factor. For 
$\Gamma_1$ it is nonzero even for $T=0$ (spontaneous transition),
whereas for $\Gamma_2$ it is zero at $T=0$. In fact, no phonon can be annihilated
if it does not exist. 

$\Gamma_1$ can be evaluated by a one-dimensional integration,
\be\label{gamma1}
\Gamma_1=8\pi n_0 a^2 \frac{v}{\eta^2} Q_\eta\left(\frac{1}{2}\beta Mc^2\right),
\ee
where $v$ is the impurity velocity, $\eta\equiv v/c$, $c$ is the velocity of sound 
and $Q_\eta$ stays for 
\be
Q_\eta(\bar\beta)=\int_1^h dx x(1-x^{-2})^2
\frac{e^{\bar\beta\left(x^2-1/x^2\right)}}
{e^{\bar\beta\left(x^2-1/x^2\right)}-1}, 
\ee
$h$ being equal to $max[1,\eta]$.
The cross section is $\sigma_1=\Gamma_1/n_0 v$.

$\Gamma_2$ can be evaluated by a similar integral, but in this case the integration
range is independent of $\eta$ if $\eta\ge1$. We have
\be\label{gamma2}
\Gamma_2=8\pi n_0 a^2 \frac{v}{\eta^2} P_\eta\left(\frac{1}{2}\beta Mc^2\right),
\ee
where 
\be
P_\eta(\bar\beta)=\int_0^h dx\frac{x(1-x^{-2})^2}
{e^{\bar\beta\left(1/x^2-x^2\right)}-1},
\ee
$h$ being equal to $min[1,\eta]$.

Using the transformation $x\rightarrow1/x$ it is simple to demonstrate that 
\be
P_{\eta\ge1}(\bar\beta)=\lim_{\eta\rightarrow\infty}\left[Q_\eta(\bar\beta)-Q_\eta(\infty)
\right],
\ee
i.e., for $\eta\gg1$, the stimulated {\it dissipative} scattering rate is equal to
the {\it cooling} scattering rate. However for a generic value of $\eta$ the
cooling scattering rate is always larger than the stimulated dissipative one, because
$Q_\eta(\bar\beta)-Q_\eta(\infty)$ is a increasing monotonic function in $\eta$.

The average impurity-phonon cross section for $(II)$ is enhanced by a 
factor $n_0$ because of the
presence of the condensate, furthermore for $\eta>1$ it depends inversely on $\eta^2$, 
it grows upon decreasing the impurity velocity or increasing the sound 
velocity. Therefore in an inhomogeneous trapped condensate the cooling scattering rate 
is lower near the borders because the condensate density $n_0$ and $c$ are smaller.

\begin{figure}
\epsfig{figure=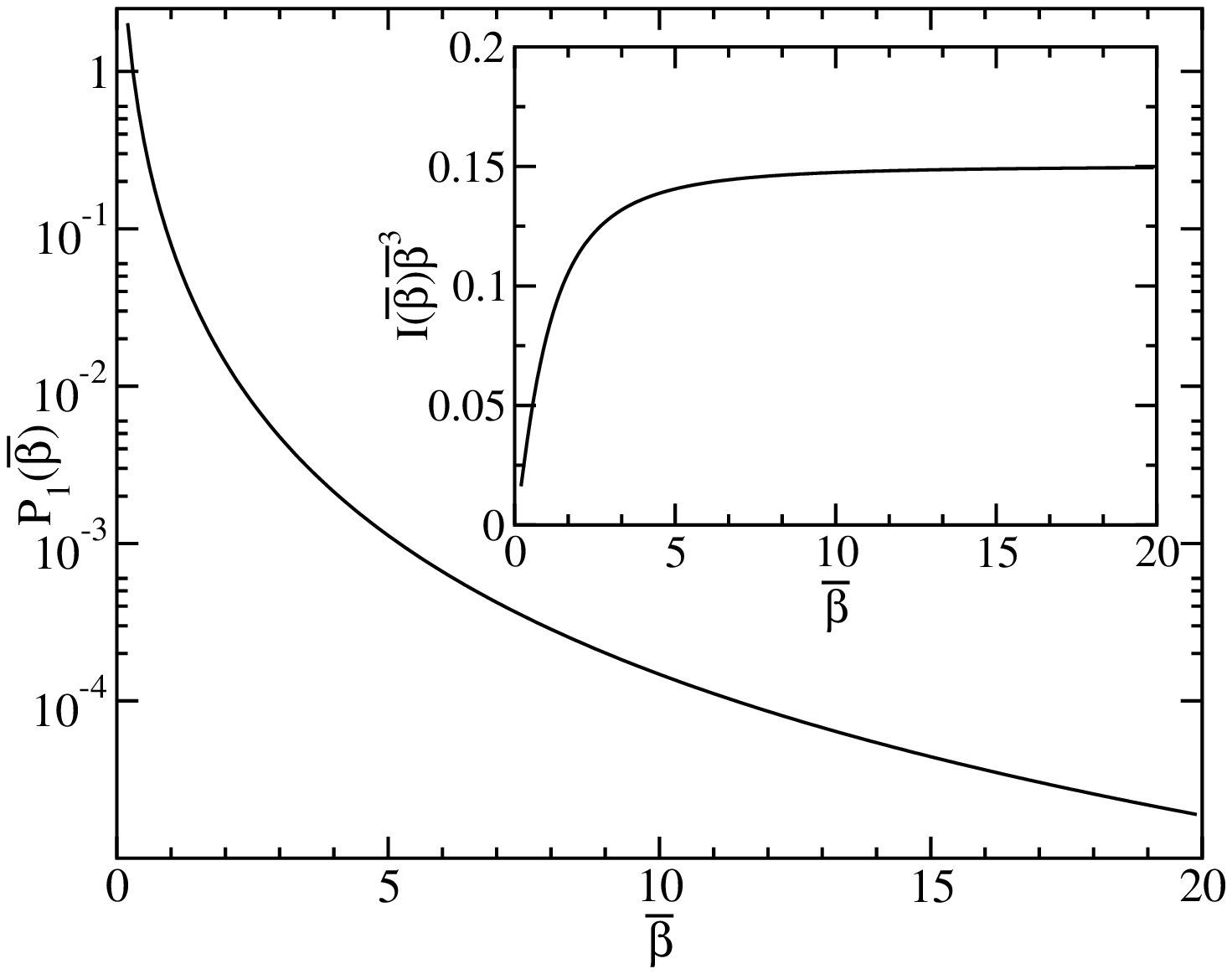,width=6.5cm,angle=-90}
\caption{The integral $P_1(\bar\beta)$ as a function of $\bar\beta$.}
\label{fig1}
\end{figure}    

In Fig.~\ref{fig1} we report $P_1(\bar\beta)$ as a function of $\bar\beta$.
Note that the integral $P_1(\beta Mc^2/2)$ drops to zero for $T\rightarrow 0$.
Since the number of thermal atoms is approximately proportional to $1/\beta^3$,
the behavior of $P_1(\bar\beta)\bar\beta^3$, as a function of $\bar\beta$, is
identical to the behavior of the scattering rate divided by the number of thermal atoms.
Such function is shown in the inset of Fig.~\ref{fig1}. We can see that for 
sufficiently high temperatures the average scattering rate of a single phonon grows
upon decreasing the temperature. This occurs because a larger percentage of phonons
fulfills the kinematic conditions.

\begin{figure}
\epsfig{figure=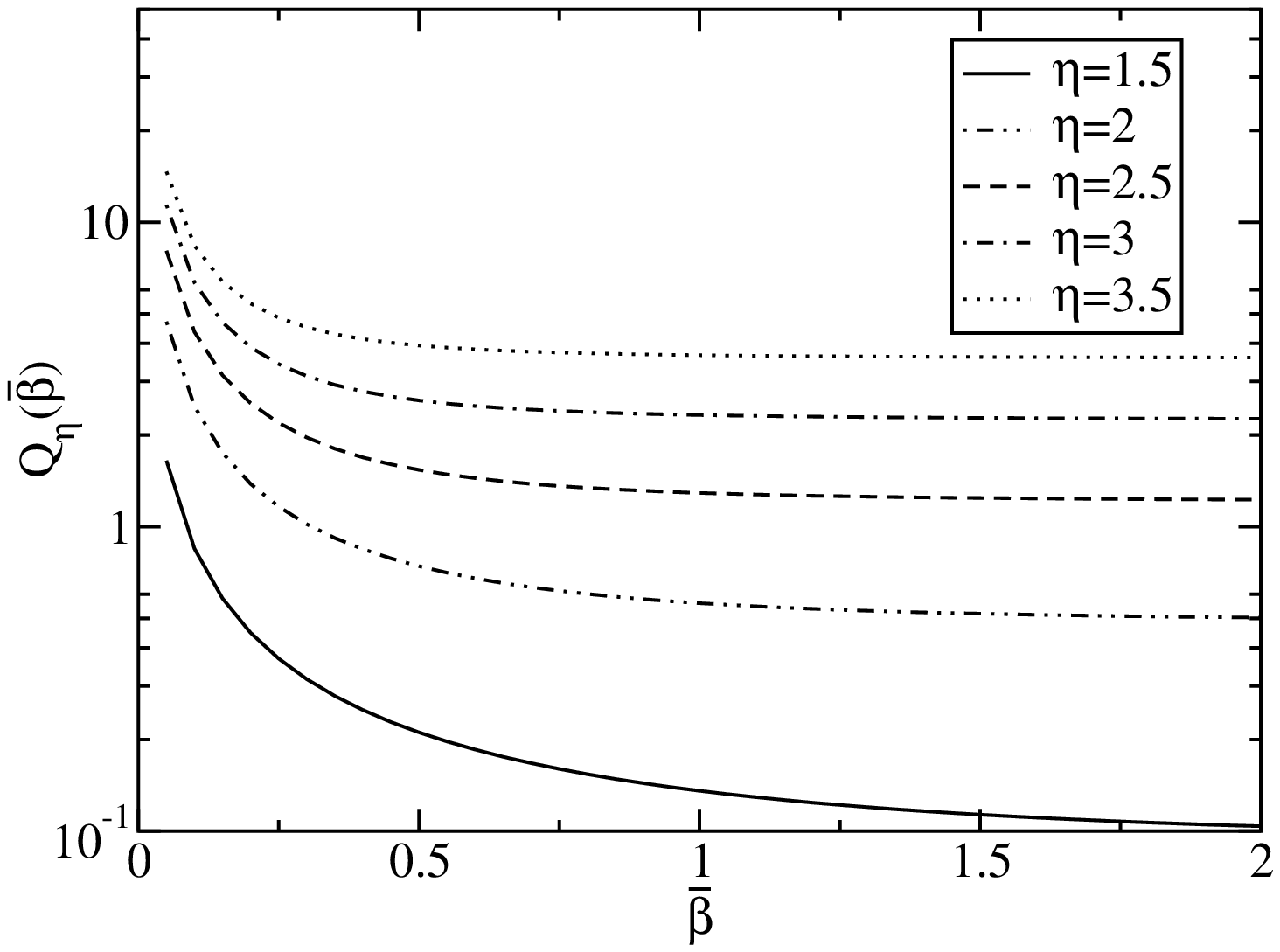,width=6.5cm,angle=-90}
\caption{$Q_\eta(\bar\beta)$ as a function of $\bar\beta$, for some values of 
$\eta$.}
\label{fig2}
\end{figure} 

In Fig.~\ref{fig2} we have plotted $Q_\eta(\bar\beta)$ as a function of $\bar\beta$ and
for some values of $\eta$. For a large $\beta$, i.e. a small temperature, 
the stimulated scattering drops to zero and only the spontaneous scattering contributes
to $Q_\eta$. For $\beta\rightarrow0$ $Q_\eta\rightarrow\infty$, because of the increase
of the phonon population. 

It is interesting to note that for values of $\eta$ near $1$ the dissipative 
scattering rate lowers and becomes zero for $\eta<1$, whereas the cooling scattering 
rate due to $(II)$ remains finite. Therefore below the Landau critical
velocity the collisions can only annihilate phonons, cooling the condensate. 
The maximum cooling rate is reached below the critical velocity.
It vanishes for an impurity velocity approaching zero, because
the required energy of the annihilated phonons increases.
Above the critical velocity the cooling collisions can prevail over the dissipative 
ones for sufficiently high temperatures. 

To refer to a realistic situation we apply
Eqs.~(\ref{gamma1}) and (\ref{gamma2}) to the data of Ref.~\onlinecite{ketterle}. 
As reported, a variation of the radial trapping frequency between $165$ and $33$ Hz 
implies a variation of the speed of sound from $1.1$ to $0.55$ cm/s. These values
require that $N_0 g_{GP}\sim 2.18\times 10^{-43} J m^3$, where $N_0$ is the number of
condensed atoms and $g_{GP}=4\pi\hbar^2 a_{1,-1}/M$, $a_{1,-1}$ being the s-wave 
scattering length of the condensate atoms. In the following we will use this value for
$N_0 g_{GP}$.
The impurity atoms are created in an untrapped hyperfine level, therefore they are 
gravitationally accelerated. For experimental reasons they are produced with an initial 
axial velocity equal to $7 mm/s$. The number of scattered atoms is obtained by counting 
the impurity atoms in a region of the time-of-flight image below the unscattered 
impurity atoms, where the impurities that have lost kinetic energy are present.
Since in this region there are Raman outcoupled thermal $m_F=0$ atoms,
the number of collided atoms in the counting region is obtained by subtracting the 
thermal background,
determined by counting a similar sized region above the unscattered impurity
atoms, where a few collision products are present.  
This background can be overvalued because some impurities, that
have acquired energy by annihilating phonons, can be in the region above the unscattered
atoms, where the presence of sole thermal atoms is supposed. In fact in 
Ref.~\onlinecite{nota} the authors conclude that, because of this overvaluation,
the data of Ref.~\onlinecite{ketterle} are not affected by finite temperature.
However, we show now that
if the velocity is sufficiently larger than the sound velocity, the atoms that scatter
annihilating phonons populate equally both regions. 

\begin{figure}
\epsfig{figure=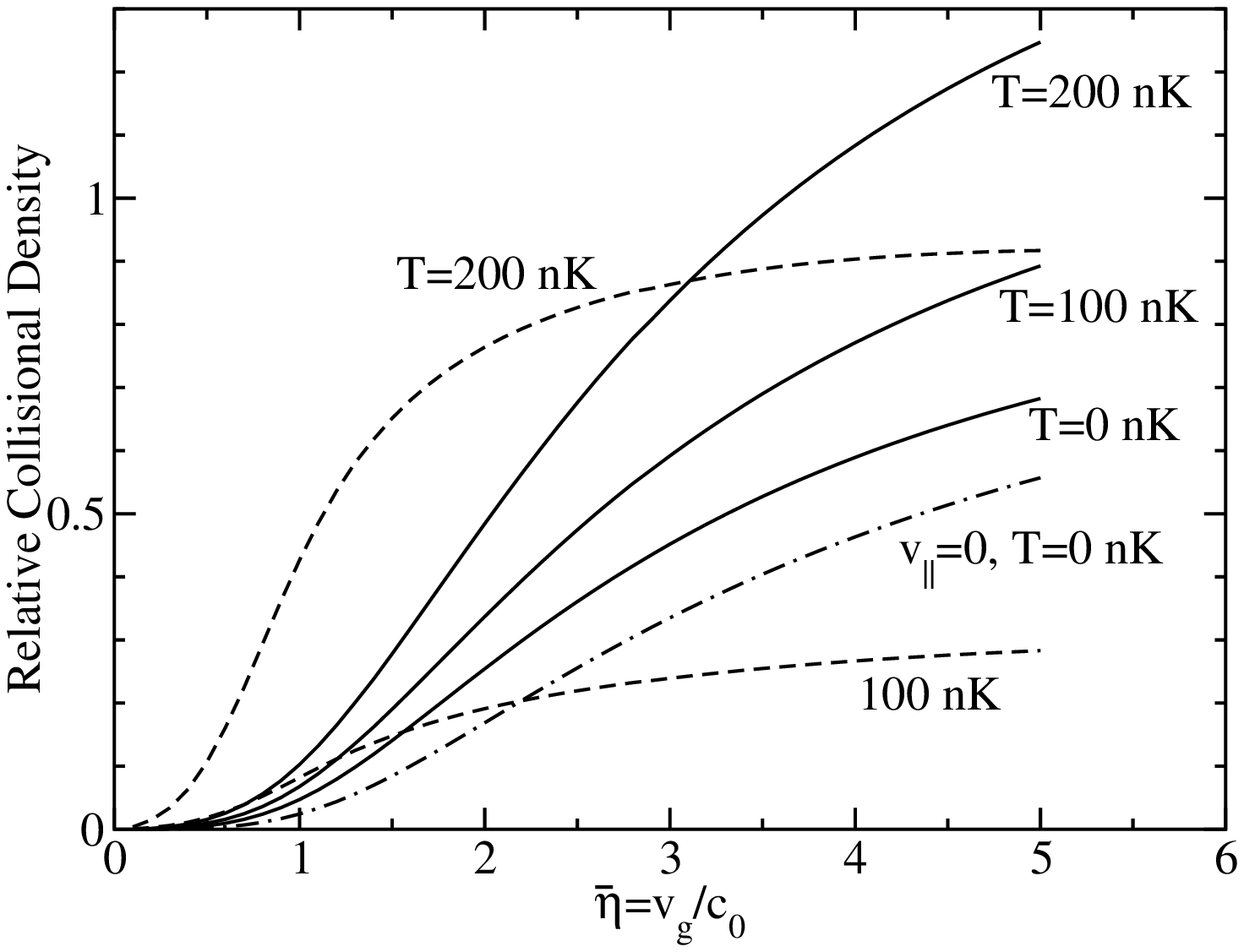,width=6.5cm,angle=-90}
\caption{Collisional density as a function of $\bar\eta$, for some values of
the temperature and for $v_\parallel=7 mm/s^{-1}$. The solid and dashed lines refer to the
channels $(I)$ and $(II)$, respectively. The dashed-dotted line is the dissipative
collisional rate for $T=0 nK$ and $v_\parallel=0$.}
\label{fig3}
\end{figure} 

\begin{figure}
\epsfig{figure=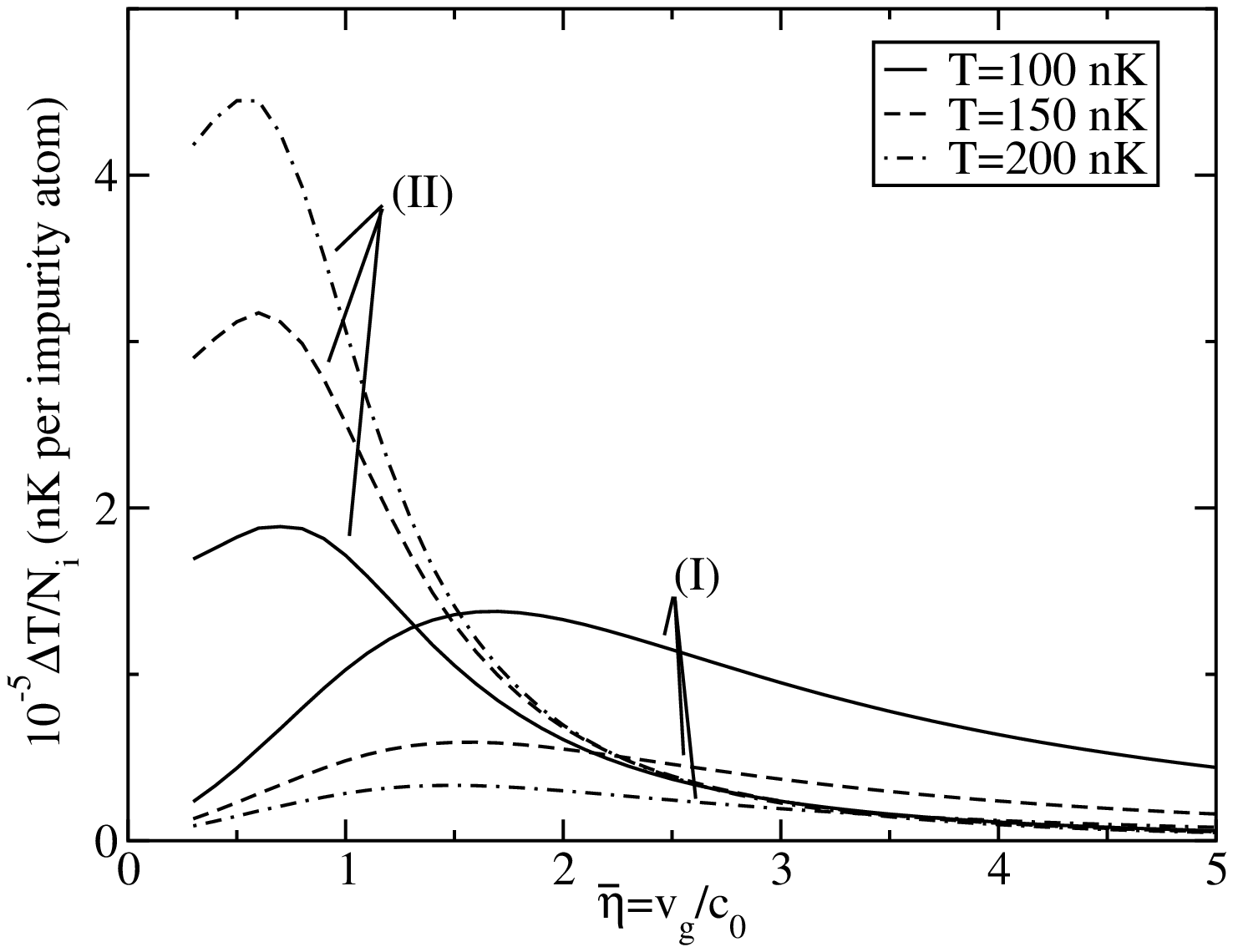,width=6.5cm,angle=-90}
\caption{$|\Delta T|$ for the two channels $(I)$ and $(II)$, as a function of 
$\bar\eta$ and for some values of $T$.} 
\label{fig4}
\end{figure} 

If the scattering creates a phonon, the impurity loses the velocity $\vec v_l$, with
$|\vec v_l|=v \cos\theta -c^2/(v\cos\theta)$.
Here $v$ is the impurity velocity and $\theta$ is the angle between that
velocity and the phonon momentum. $\theta$ goes from $0$ to $\arccos(c/v)$, therefore
the lost velocity has a direction around the impurity one. Because of this 
directionality
the dissipative scattering products move in the region of the time-of-flight image 
below the unscattered impurity atoms. 
If the scattering annihilates a phonon, the impurity increases the velocity by
$\vec v_a$, with $|\vec v_a|=c^2/(v\cos\theta)-v \cos\theta$,
where $\theta$ ranges between $\arccos(c/v)$ and $\pi/2$. 
If $v/c\gg1$ the acquired velocity is near orthogonal to the impurity velocity, 
i.e. the cooling scattering products are not confined within a narrow cone as
the particles that collide creating phonons. Therefore we expect
that the products of $(II)$ are equally distributed in both the measurement
regions~\cite{nota2} and thus their contribution is negligible because of the 
background subtraction method.
For example, with $v/c=2$ and $v/c=5$, $\theta$ ranges, respectively, between 
$60^o$ and $90^o$ and between $80^o$ and $90^o$ (the velocities are distributed near the
surface of an open umbrella); furthermore 
its mean value approaches $90^o$ upon increasing the temperature.

The problem is complicated both by the inhomogeneity of the condensate and
the continuous variation of the impurity direction of motion during the free fall. 
Therefore to understand or to evaluate the exact contribution of the two collisions
channels to the experimental data is not a trivial task. We have calculated only the
overall collisional density for each channel.

If the initial axial velocity is small, an impurity produced at the point 
$(x,y,z)$ has the following probability to collide dissipatively~\cite{ketterle}
\be
p(x,y,z)=\int_z^{e(x,y)}dz_1n(x,y,z_1)\sigma_1(\eta,c(x,y,z_1)),
\ee
where $\eta$ is determined by the local condensate density and the downward impurity 
velocity, $n$ is the condensate density 
and $e(x,y)=\sqrt{2\mu/m\omega_z^2-(\omega_x^2 x^2+\omega_y^2y^2)/\omega_z^2}$.
$c(\vec x)$ is $c_0 \sqrt{n(\vec x)/n(0)}$, $c_0$ being the speed of sound at the
center of the condensate.
Note that for $1/\beta=0$ the $\sigma_1$ 
function depends only on $\eta$, as considered in Ref.~\onlinecite{ketterle}.
As a first approximation, we have taken into account also the axial velocity 
using the total velocity, and not only the vertical component, to evaluate $\eta$.
We have also replaced $dz_1$ with $dz_1(v_\parallel^2+v_z^2)^{1/2}/v_z$, where 
$v_\parallel$ and $v_z$ are the axial and vertical velocities, respectively.
We have supposed that the condensate is nearly homogeneous in the axial direction, 
so the variation of sound velocity and of $e(x,y)$ in that direction is negligible.

The total number of dissipatively scattered atoms is
$C_\beta(\bar\eta)=\int d\vec r n_I(\vec r) p(\vec r)$,
where $n_I(\vec r)$ is the initial impurity density and $\bar\eta\equiv v_g/c_0$,
$v_g\equiv\sqrt{2gz_c}$ being the average vertical velocity and $z_c$ the Thomas-Fermi 
radius in the z direction. We can assume 
$n_I\propto n$~\cite{ketterle}. In Fig.~\ref{fig3} we have reported (solid lines)
$C_\beta(\bar\eta)/C_\infty$ as a function of $\bar\eta$ and for some values
of the temperature $T$. $C_\infty$ is the collisional
density for high velocities and zero temperature.
We have reported also the collisional density for the channel $(II)$ (dashed line), 
calculated in a similar fashion. The dashed-dotted line refers to $(I)$
and is evaluated for $T=0$ and $v_\parallel=0$. We can see that the contribution of the 
axial velocity is important also for high $\bar\eta$. This is true especially for
$(II)$, because the scattering rate is higher for small velocities and,
therefore, we have an important contribution at the beginning of the free fall.
In particular, for high $\bar\eta$ the horizontal velocity decreases the scattering
rate, because it is higher than $v_L$. In fact, the cooling scattering rate has a 
maximum peak below $v_L$.
For $\bar\eta=5$ the temperature extracted from the time-of-flight pictures is
about $100 nK$~\cite{kettco}. For this value the dissipative scattering rate is enhanced
by about $30\%$. 

For low $\bar\eta$ the cooling scattering rate is higher than the dissipative one, 
suggesting that the overall effect of the collisions is the cooling of the condensate. 
We have calculated for each channel the average energy loss or gain of
the trapped system per impurity atom. The temperature growth per impurity atom is
$\Delta T=\Delta E/C_N(T)$,
where $\Delta E$ is the energy increase per impurity atom and $C_{N}(T)$ is the 
heat capacity, which for $N$ interacting trapped atoms is~\cite{stringari}
\be
C_N=\left(\frac{\partial E}{\partial T}\right)_N=
N k_B\left[12\frac{\zeta(4)}{\zeta(3)}t^3+1.2\gamma\frac{t^2(5-8t^3)}{(1-t^3)^{3/5}}\right].
\ee
$k_B$ is the Boltzmann constant, $\zeta$ the Riemann function, $t$ is $T$ divided by
the critical temperature, and
$\gamma=1.57(N^{1/6}a_{1,-1}/a_{ho})^{2/5}$,
$a_{ho}$ being the geometric average of the three harmonic oscillator lengths.
$C_N$ has a very weak dependence upon $N$ for $t\ll1$, we have used the value 
$N=2\times10^7$. In Fig.~\ref{fig4} we have reported $|\Delta E|$
as a function of $\bar\eta$ for each channel and for some values of $T$. 
For small $\bar\eta$ and high temperatures
the cooling rate prevails over the heating one, i.e, the condensate cools. While $T$ 
decreases,
the heating rate grows and the mimimum theoretic temperature is reached when the two
contributions balance. For a tight trap the typical temperature of a condensate is 
$\sim200 nK$. With $\bar\eta=1.5$, $T=200 nK$ and $5\times 10^6$ impurity atoms
the temperature decrease is about $25 nK$, if we neglect the variation of $C_N$ and suppose
that the thermalization velocity is larger than the cooling one. For $\bar\eta=0.75$
$(\nu_\bot\sim300 Hz)$
and $T=200 nK$, $|\Delta T|$ is $\sim100 nK$, which is a remarkable value.

In conclusion, we have considered the effects of finite temperature on the scattering of
impurity atoms and we have shown that there are two important collision channels.
In $(I)$ the collisions are dissipative and the impurities lose energy 
heating the condensate, in $(II)$ the impurities acquire energy cooling the 
condensate. 
Below the Landau velocity only $(II)$ has a non zero scattering rate.
We have calculated the scattering rate for each channel with reference to the
experimental configuration of Ref.~\onlinecite{ketterle} and we have shown that
the temperature may have a visible effect upon the experimental data. We have
evaluated the energy loss or gain of the trapped atoms and demonstrated that
the overall scattering process cools the condensate for tight traps and sufficiently
high temperatures. This effect can become much more efficient using different 
experimental configurations~\cite{future}.

I am grateful to Professor F. T. Arecchi for helpful discussions, A. P. Chikkatur
for useful exchange of information and Professor W. Ketterle for valuable comments.
This work is partly supported by CNR contract n. CT98.00154.


\begin{thebibliography}{20}
\bibitem{nota2} For large $\bar\eta$ the average velocity of the unscattered impurities 
is nearly orthogonal to the time-of-flight image plane.
\end{thebibliography}
\end{document}